 \documentclass[prl,amsmath,amssymb,twocolumn, showpacs, superscriptaddress,10pt]{revtex4-2}

\usepackage{amsmath}
\usepackage{hyperref}
\usepackage{graphicx}
\usepackage{amsfonts}
\usepackage{amsthm}
\usepackage{cases}
\usepackage{bm}

\usepackage{fixmath}
\newcommand{\vect}[1]{\mathbold {#1}} 

\usepackage{enumitem} 

\usepackage[normalem]{ulem}

\usepackage{color}
\definecolor{Blue}{rgb}{0.00, 0.00, 0.80}
\definecolor{Red}{rgb}{0.80, 0.00, 0.00}
\definecolor{Green}{rgb}{0.00, 0.50, 0.00}

\hypersetup{
    colorlinks=true,       
    linkcolor=red,          
    citecolor=blue,        
    filecolor=magenta,      
    urlcolor=cyan           
}

\newcommand{\antiquad}{\!\!\!\!\!\!\!\!}
\newcommand{\nn}{\nonumber}
\newcommand{\be}{\begin{equation}}
\newcommand{\ee}{\end{equation}}
\newcommand{\bea}{\begin{eqnarray}}
\newcommand{\eea}{\end{eqnarray}}

\newcommand{\beq}{\begin{equation}}
\newcommand{\eeq}{\end{equation}}
\newcommand{\beqn}{\begin{eqnarray}}
\newcommand{\eeqn}{\end{eqnarray}}

\begin{document}

\title{Nonequilibrium steady state of trapped active particles}

\author{Naftali R. Smith}
\email{naftalismith@gmail.com}
\affiliation{Department of Environmental Physics, Blaustein Institutes for Desert Research, Ben-Gurion University of the Negev, Sede Boqer Campus, 8499000, Israel}

\begin{abstract}

We consider an overdamped particle with a general physical mechanism that creates noisy active movement (e.g., a run-and-tumble particle or active Brownian particle etc.), that is confined by an external potential.
Focusing on the limit in which the correlation time $\tau$ of the active noise is small, we find the nonequilibrium steady-state distribution $P_{\text{st}}\left(\vect{X}\right)$ of the particle's position $\vect{X}$.
While typical fluctuations of $\vect{X}$ follow a Boltzmann distribution with an effective temperature that is not difficult to find, the tails of $P_{\text{st}}\left(\vect{X}\right)$ deviate from a Boltzmann behavior: In the limit $\tau \to 0$, they scale as $P_{\text{st}}\left(\vect{X}\right)\sim e^{-s\left(\vect{X}\right)/\tau}$.
We calculate the large-deviation function $s\left(\vect{X}\right)$ exactly for arbitrary trapping potential and active noise in dimension $d=1$, by relating it to the rate function that describes large deviations of the position of the same active particle in absence of an external potential at long times. We then extend our results to $d>1$ assuming rotational symmetry.

\end{abstract}

\maketitle

\textit{Background} --- 
Active particles propel themselves by pumping energy from their environment \cite{Romanczuk,soft,BechingerRev, FodorEtAl15, Fodor16, Needleman17, Ramaswamy2017,Marchetti2017,Schweitzer, FJC22}. Active motion is not symmetric under time reversal, thus providing examples of systems that are out of thermal equilibrium even at the single-particle level.
Natural examples of active systems are found on a wide variety scales, 
including molecular motors \cite{BHG06, Toyota11,  Stuhrmann12, Mizuno07},
living cells and/or bacteria  \cite{Berg2004,Wilhelm08, Cates2012, Ahmed15, BeerEtAl20, Breoni22, Nishiguchi23}, 
birds \cite{flocking1, flocking2} and fish \cite{Vicsek,fish}. Inspired by these natural examples, many artificial active systems, consisting of colloids, Janus particles (objects composed of two or more parts that differ in their physical and/or chemical properties \cite{GLA07}) or self-propelled robots have been fabricated and studied experimentally \cite{BechingerRev, Hagen2014,Takatori2016,Deblais2018,Dauchot2019, gran2, PBDN21, BCASL22, MolodtsovaEtAl23, PPSK23, Nishiguchi23}.

Active systems can display remarkable, nonequilibrium collective behaviors such as motility-induced phase separation \cite{separation1, separation2, separation3}, clustering and/or flocking \cite{cluster1,cluster2,evans} and surprising boundary-related effects \cite{Kardar2015}. 
However, even a \emph{single} active particle can display behaviors that are qualitatively different to its passive (Brownian) counterpart. 
For instance, an active particle tends to aggregate near the boundaries of a confining region, unlike a Brownian particle which occupies the confining region homogeneously \cite{Wensink2008, Elgeti2009, Li2009b,TC08, Kaiser2012, Elgeti2013, Hennes2014, Solon2015, Takatori2016, Li2017, Razin2017, Dauchot2019, Pototsky2012}.
More generally, an active particle trapped by an external potential reaches a non-Boltzmann, nonequilibrium steady state \cite{Solon2015,Pototsky2012,TC08, ABP2019,Dhar_2019,Malakar20, Franosch2016,Das2018,Caprini2019,Sevilla2019, Hagen2014,Takatori2016,Deblais2018,Dauchot2019, SBS21}, and its first-passage properties in general deviate from the Arrhenius law \cite{MalakarEtAl18,ABP2018, Singh2019}.

Calculating the statistical properties of this nonequilibrium steady state is in general a difficult task, which has been achieved only in special individual cases, while a general paradigm is lacking. In particular, while typical fluctuations are sometimes easier to understand as they behave similarly to thermal (passive) systems, the large-deviations regime, which often includes additional signatures of activity, is far less understood.

\textit{Model and summary of main results} ---
Several theoretical models have been proposed, that mimic natural or artificial active systems.
A generic theoretical model for overdamped active particles can be written in the form of the stochastic differential equation
\be
\label{LangevinJustNoise}
\vect{\dot{x}} = \vect{\sigma}(t),
\ee
where $\vect{\sigma}(t)$ represents some (random) noise term that originates in the self-propulsion of the particle and/or its interaction with its environment. 
We denote the correlation time of the noise by  $\tau$.
The broad class of models that can be written in the form \eqref{LangevinJustNoise} includes (see precise definitions below) the active Ornstein-Uhlenbeck particle (AOUP) \cite{FodorEtAl15}, the run-and-tumble particle (RTP) \cite{TC08}, the active Brownian particle (ABP) \cite{BechingerRev} and passive Brownian motion as particular cases.
A natural way to quantify fluctuations in such models is through the distribution of the position $\vect{x} =\vect{x}(t) $ of the particle at time $t$, given that it begins at the origin $\vect{x}(t=0) = 0$. At long times, many models (including all of the examples mentioned above) show a universal diffusive behavior: Typical fluctuations of $\vect{x}$ are described by a Gaussian distribution whose variance grows linearly in time,
\be
\label{GaussianFree}
P_{\text{free}}\left(\vect{x},t\right)\simeq\frac{1}{\sqrt{4\pi D_{\text{eff}}t}}e^{-x^{2}/4D_{\text{eff}}t},
\ee
with some effective diffusion coefficient $D_{\text{eff}}$. 
%
Nevertheless, signatures of activity of the process $\vect{\sigma}$ remain at arbitrarily long times in the \emph{tails} of $P_{\text{free}}\left(\vect{x},t\right)$. For many models (including, again, all the examples given above), these tails  follow a large-deviations principle (LDP) \cite{Touchette2018, ABP2019, SBS20, Dean21}
\be
\label{LDPfree}
P_{\text{free}}\left(\vect{x},t\right)\sim e^{-(t/\tau)\Phi\left(\vect{x}/t\right)}
\ee
with a process-dependent convex ``rate function'' 
$\Phi\left(\vect{z}\right) \ge 0$ 
that has a unique minimum $\vect{z}=\vect{z}^*$ at which 
$\Phi\left(\vect{z}^*\right) = 0$.
For simplicity, we assume $\vect{z}^*=0$ \cite{footnote:drift}.
This LDP \eqref{LDPfree} is valid in the limit $t \to \infty$ with $\vect{x}/t$ and $\tau$ fixed or
equivalently, in the limit $\tau \to 0$  with $\vect{x}$ and $t$ fixed.
For the examples mentioned above, $\Phi(\vect{z})$ is known, see Table \ref{table:PhiLambdaMu}.
$\Phi(\vect{z})$ is generically quadratic around its minimum, thus providing a smooth matching with the Gaussian, typical-fluctuations regime \eqref{GaussianFree}.

\begin{table}


 \begin{tabular}{ | l | l | l | l |}
    \hline
Process & $\Phi(z)$ & $\lambda(k)$ & $\mu(u)$ \\ \hline
AOUP &  $z^2/2$ &  $k^2/2$ & $-2u$  \\ \hline
Symmetric RTP & $1-\sqrt{1-z^2}$ &  $\sqrt{k^{2}+1}-1$ & $2u/ \! \left(u^{2}-1\right)$  \\ \hline
Asymmetric RTP & --- &  Eq.~\eqref{lambdakAsymmtricRTP} & $\frac{1}{1+u}-\frac{\alpha}{\alpha-u}$  \\ \hline
PRW & see \cite{SM} &  $\ln\left(\cosh k\right)$ & ---  \\ \hline
ABP & --- &  $-a_{0}\left(-2k\right) \! / 4$ & ---  \\ \hline
    \hline
    \end{tabular}
    \caption{Rate functions $\Phi(z)$ and their corresponding SCGFs $\lambda(k)$ that describe fluctuations of a free active particle for several different active noises, together with the associated functions $\mu(u)$ that yield the SSD and MFPT for a trapped active particle in the limit of short correlation time of the active noise, see Eq.~\eqref{sXsol}. The result for the symmetric RTP is valid in $d=1$ and $d=2$. 
Results for the ABP are given in terms of the smallest eigenvalue $a_0(q)$ of the Mathieu equation \eqref{Mathieu}. Wherever $\Phi$ or $\mu$ are marked by a ---, we obtain them from $\lambda(k)$ numerically.}
\label{table:PhiLambdaMu}
\end{table}

Another natural physical setting, often encountered in experiments \cite{Hagen2014,Takatori2016,Deblais2018,Dauchot2019}, is that of an active particle trapped by an external potential. Then the equation of motion becomes
\be
\label{Langevin}
\dot{\vect{x}}=\vect{F}\left(\vect{x}\right)+\vect{\sigma}\left(t\right),
\ee
where $\vect{F}\left(\vect{x}\right)= -\nabla U\left(\vect{x}\right)$ is the force exerted by the external trapping potential $U$ which is assumed to have a unique global minimum at $\vect{x}=0$.
The trapping potential alters the behavior considerably, compared to a free particle: At long times, the particle's position is expected to converge to a nonequilibrium steady state distribution (SSD)
$P_{\text{st}}\left(\vect{X}\right)$ \cite{Solon2015,Pototsky2012,TC08, ABP2019,Dhar_2019,Malakar20, Franosch2016,Das2018,Caprini2019,Sevilla2019, Hagen2014,Takatori2016,Deblais2018,Dauchot2019, SBS21}.
If $\vect{\sigma}$ is a passive noise, then it can be described by a white, Gaussian noise with
$\left\langle \vect{\sigma}\left(t\right)\right\rangle =0$
and
$\left\langle \sigma_{i}\left(t\right)\sigma_{j}\left(t'\right)\right\rangle =\sqrt{2D} \, \delta_{ij}\delta\left(t-t'\right)$,
 where $D$ is the diffusion coefficient and angular brackets denote ensemble averaging. 
In this case, the process is in thermal equilibrium, and 
$P_{\text{st}}\left(\vect{X}\right) \! \propto \! e^{-U\left(\vect{X}\right)/D}$
 is given by the Boltzmann distribution.
Here $D = k_B T$,  where $k_B$ and $T$ are Boltzmann's constant and the temperature, respectively.
However, if $\vect{\sigma}$ is an active noise, calculating $P_{\text{st}}\left(\vect{X}\right)$ becomes a major challenge.

A related question is at what time $\tau_\vect{X}$ will a particle, starting from $\vect{x}=0$, first reach position $\vect{X}$? For passive (Brownian) particles, the answer to this question is known at $U\left(\vect{X}\right) \gg D$: The first passage time $\tau_\vect{X}$ follows an exponential distribution whose mean is given, in the leading order, by the Arrhenius law (Kramers' formula) \cite{Kramers40, Hanggi90, Melnikov91}
$\left\langle \tau_{\vect{X}}\right\rangle \sim1/P_{\text{st}}\left(\vect{X}\right)\sim e^{U\left(\vect{X}\right)/D}$.

One active system in which the SSD is known exactly is the RTP in dimension $d=1$, in an arbitrary potential $U(x)$ \cite{q-optics1,q-optics2,q-optics3,q-optics4, VBH84, colored, Kardar2015, Dhar_2019}, as we recall in the supplemental material \cite{SM}.
For the particular case of a harmonic trapping potential, these results in $d=1$ were recently extended to the case of a general noise term that undergoes periodic or Poissonian resetting \cite{GMS23}.
$P_{\text{st}}\left(\vect{X}\right)$ and closely related quantities have been recently found exactly for several variants of the RTP model in $d=2$ or higher \cite{Frydel22,SLMS22}
and studied also for the ABP \cite{ABP2019, Malakar20, CD21, CSLW22, NG22, BCASL22, CF22} for the particular case of a harmonic trapping potential $U(\vect{x}) \propto x^2$.

The goal of this Letter is to calculate the SSD 
and mean first passage times (MFPTs) for a broad class of potentials $U(\vect{x})$ and active noises $\vect{\sigma}(t)$, in the small-$\tau$ limit \cite{footnote:smallTau},
i.e., when the typical timescale of the active noise is much shorter than the timescale associated with the external force $\vect{F}$.
We find that typical fluctuations obey a Boltzmann distribution
$P_{\text{st}}\left(\vect{X}\right)\sim e^{-U\left(\vect{X}\right)/k_B T_{\text{eff}}}$
where $T_{\text{eff}}$ is an effective temperature that we calculate.
However, the \emph{tails} of the SSD behave very differently. Using tools from large-deviations theory, we develop a general framework for calculating these tails, and find that (at $\tau\to0$) they satisfy an LDP
$P_{\textrm{st}}\left(\vect{X}\right)\sim e^{ - s\left(\vect{X}\right)/\tau}$.
In $d=1$, we calculate the large-deviation function (LDF) $s\left(X\right)$ exactly by relating it to the rate function $\Phi(z)$, see Eq.~\eqref{sXsol} below. Thus, we uncover a remarkable connection between the statistics of active particles in the free and trapped cases.
Our results extend immediately to $d>1$, assuming rotational symmetry.
Finally, we apply our general formalism to several particular models of active particles, obtaining the results presented in Table \ref{table:PhiLambdaMu}.

\textit{Theoretical framework} ---
For simplicity, we begin from the case $d=1$.
In the small-$\tau$ limit, we exploit the timescale separation between the timescale $\tau$ of the noise and the the relaxation time of the particle in the potential in the absence of noise.
In this limit, we coarse grain the noise by averaging it over time windows of intermediate size (much longer than $\tau$, but much shorter than the timescales of $U(x)$).
By applying the LDP \eqref{LDPfree} to each of these windows, the probability of a coarse-grained noise history $\bar{\sigma}(t)$ is given by \cite{HT09, Harris15, Jack19, JH20, AgranovBunin21, SmithFarago22, BTZ23}
$P\left[\bar{\sigma}\left(t\right)\right]\sim e^{-s\left[\bar{\sigma}\left(t\right)\right] / \tau}$
where 
$s\left[\bar{\sigma}\left(t\right)\right]=\int\Phi\left(\bar{\sigma}\left(t\right)\right)dt$
is interpreted as the action functional of an underlying Hamiltonian system, as we show below.

Replacing the noise term in \eqref{Langevin} by its coarse-grained counterpart, we obtain the coarse-grained Langevin equation
\be
\label{LangevinCG}
\dot{x}=F\left(x\right)+\bar{\sigma}\left(t\right).
\ee
We now apply the optimal fluctuation method (OFM) to Eq.~\eqref{LangevinCG}. This involves using a saddle-point approximation on the path integral that corresponds to the stochastic dynamics \eqref{LangevinCG}, resulting in a minimization problem for the ``optimal'' (i.e., most probable) history of the system  conditioned on observing the event of interest.
Since we are interested in the SSD of the particle's position, we let the system evolve from time $t=-\infty$ to time $t=0$, at which the position is measured, $X = x(t=0)$.
To calculate the probability of a (coarse-grained) history $x(t)$,  one simply eliminates $\bar{\sigma}$ from \eqref{LangevinCG}, leading to
$P\left[x\left(t\right)\right]\sim e^{-s\left[x\left(t\right)\right] / \tau}$
with 
\be
\label{action}
s\left[x\left(t\right)\right]=\int_{-\infty}^{0}\Phi\left(\dot{x}-F\left(x\right)\right)dt \, .
\ee
At $\tau \to 0$, the dominant contribution to $P_{\text{st}}(X)$ comes from the (coarse-grained) history $x(t)$ that minimizes 
the action \eqref{action}, under the constraints
$ x\left(t\to-\infty\right)=0$ 
and $ x\left(t=0\right) = X$.

Since the Lagrangian $L=\Phi\left(\dot{x}-F\left(x\right)\right)$ in \eqref{action} does not explicitly depend on $t$, the Hamiltonian $H$ is conserved. To calculate it, we first find the ``conjugate momentum'' to $x$,
$p=\partial L/\partial\dot{x}=\Phi'\left(\dot{x}-F\left(x\right)\right)$.
Then
\bea
\antiquad H=\dot{x}p-L&=&F\left(x\right)p + \left[\dot{x}-F\left(x\right)\right]\Phi'\left(\dot{x}-F\left(x\right)\right)\nn\\
&-&\Phi\left(\dot{x}-F\left(x\right)\right)=\lambda\left(p\right)+F\left(x\right)p = E,
\eea
where $\lambda(k)$ is the long-time scaled cumulant generating function (SCGF) of the position of a free particle, which is related to $\Phi(z)$ via Legendre-Fenchel transform \cite{Touchette2018, footnote:LegendreFenchel},
\be
\Phi\left(z\right)=\sup_{k\in\mathbb{R}}\left[kz-\lambda\left(k\right)\right],
\ee
and $E$ is a constant.
From the boundary condition at $t \to -\infty$ (together with $F(0) = 0$), we find that $E=0$, i.e., we obtain
$F\left(x\right)=-\lambda\left(p\right)/p$,
or
$p = \mu(F(x))$, where $k = \mu(u)$ is the solution to the equation $u = -\lambda(k) / k$.
$P_{\textrm{st}}\left(X\right)$ is found by evaluating the action \eqref{action} on the optimal history $x(t)$, which, using our expression for $p(x)$ together with $E=0$, simplifies to
\be
\label{sXsol}
s(X)=\int_{0}^{X}p\left(x\right)dx=\int_{0}^{X}\mu\left(F\left(x\right)\right)dx
\ee
so (in the leading order)
$P_{\textrm{st}}\left(X\right)\sim e^{-s\left(X\right)/\tau}$.
Eq.~\eqref{sXsol} is a central result of this Letter. As an immediate consequence, the MFPT to reach position $X$ is given (in the small $\tau$ limit) by
$\left\langle \tau_X\right\rangle\sim1/P_{\text{st}}\left(X\right)\sim e^{s\left(X\right)/\tau}$.
These expressions for $P_{\textrm{st}}\left(X\right)$ and $\left\langle \tau_X\right\rangle$ are analogs of the Boltzmann distribution and the Arrhenius law, respectively, for active systems in the limit $\tau\to0$.

\begin{figure}[t]
\includegraphics[width=0.98\linewidth,clip=]{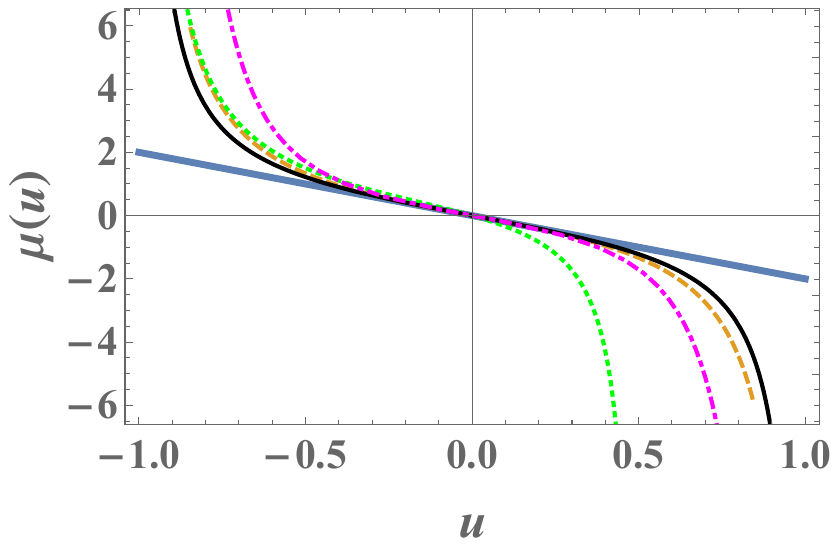}
\caption{The function $\mu(u)$ that describes the SSD and MFPT of an active particle in an external potential in the limit where the correlation time $\tau$ of the active noise is short, see Eq.~\eqref{sXsol}. 
The thick solid, dashed, dotted, thin solid and dot-dashed lines correspond to the AOUP, 
the symmetric RTP, the asymmetric RTP with $\alpha=1/2$, the PRW and the ABP, respectively.}
\label{figMuOfu}
\end{figure}

\textit{Applications and extension to $d>1$} ---
We now calculate $\mu(u)$ for several particular models of active particles.
The model whose analysis turns out to be simplest is the AOUP \cite{FodorEtAl15}, which we define via
$\sigma(t)=\Sigma(t/\tau)$
with
$\dot{\Sigma}=-\Sigma+\xi\left(t\right)$,
where $\xi\left(t\right)$ is white Gaussian noise with $\left\langle \xi\left(t\right)\right\rangle =0$ and 
$\left\langle \xi\left(t\right)\xi\left(t'\right)\right\rangle =\delta\left(t-t'\right)$.
For the AOUP,
$\Phi(z) = z^2 / 2$ \cite{Touchette2018}, coinciding with that of a passive (Brownian) particle.
The Legendre-Fenchel transform of $\Phi$ is 
$\lambda(k) = k^2 / 2$, so the inverse function of $-\lambda(k)/k$ is
$\mu(u) = -2u$.
Using this in \eqref{sXsol}, we find 
$P_{\text{st}}\left(X\right)\sim e^{-\frac{2}{\tau}\left[U\left(X\right)-U\left(0\right)\right]}$,
which is simply a Bolztmann distribution with effective temperature $k_B T_\text{eff} = \tau/2$.

A useful benchmark for our formalism is the RTP \cite{TC08}, for which $\sigma(t)$ is a telegraphic (dichotomous) noise of unit amplitude, $\sigma(t) = \pm 1$, switching sign at a constant rate $\tau^{-1}$.
The free rate function is \cite{MeersonZilber18, Dean21}
$\Phi(z) = 1-\sqrt{1-z^2}$
and its Legendre-Fenchel transform is
$\lambda\left(k\right)=\sqrt{k^{2}+1}-1$,
leading to 
$\mu\left(u\right) = 2u/ \! \left(u^{2}-1\right)$.
Indeed, using this in \eqref{sXsol}, we reproduce the leading-order term of the exact (known) result for the SSD \cite{q-optics1,q-optics2,q-optics3,q-optics4, VBH84, colored, Kardar2015, Dhar_2019}, see \cite{SM}. 

Now we find 
$\mu(u)$ for several models for which the exact SSD is not known.
Consider an asymmetric RTP in $d=1$ with different left and right speeds, so that $\sigma(t)$ takes the values $1$ and $-\alpha$, with transition rates $1/\tau$ and $\alpha/\tau$ (from $\sigma = 1$ to $\sigma = -\alpha$ and vice versa, respectively). It is easy to show that
$\left\langle \sigma\left(t\right)\right\rangle =0$ \cite{SM}.
The corresponding SCGF is \cite{MeersonZilber18,SM}
{\small
\be
\label{lambdakAsymmtricRTP}
\lambda \! \left(k\right) \! = \!  \frac{-\alpha\left(k+1\right)+ \!  \sqrt{ \! \left(\alpha+1\right)\left[\alpha\left(k+1\right)^{2}+\left(k-1\right)^{2}\right]}+k\!-\!1}{2},
\ee
}
leading to $\mu\left(u\right)=1/\left(1+u\right)-\alpha/\left(\alpha-u\right)$.

One can also consider an RTP with general distributions of waiting times between tumbling events, other than the exponential distribution that corresponds to a constant tumbling rate. Thus, our analysis extends beyond Markov processes. For instance, a Pearson random walk (PRW) \cite{Pearson1905, Kiefer84, Garcia12} in $d=1$ tumbles with probability $1/2$ at each of the times $t=\tau,2\tau,3\tau, \dots$. For the PRW, $\lambda\left(k\right)=\ln\left(\cosh k\right)$ \cite{SM}, so $\mu(u)$ is the inverse of the function $-\ln\left(\cosh k\right)/k$. We obtain the asymptotic behaviors 
\be
\label{muOfuPRW}
\mu\left(u\right)=\begin{cases}
-2u-\frac{4u^{3}}{3}+\dots, & u\to0,\\[2mm]
-\text{sgn}\left(u\right)\frac{\ln2}{1-\left|u\right|}+\dots, & |u|\to1
\end{cases}
\ee
of this $\mu(u)$ in \cite{SM}.

Our approach extends naturally to dimensions $d>1$. The action
\eqref{action} 
is simply replaced by
$s\left[\vect{x}\left(t\right)\right]=\int_{-\infty}^{0}\Phi\left(\dot{\vect{x}}-\vect{F}\left(\vect{x}\right)\right)dt$.
It is particularly simple to consider the rotationally-symmetric case, in which
$U(\vect{r}) = U(r)$
is a central potential, and the statistics of the noise $\vect{\sigma}$ are rotationally invariant, leading to a rotationally-invariant rate function
$\Phi(\vect{z}) = \Phi(z)$.
Then $P_{\text{st}}(\vect{X})$ is rotationally-symmetric too.
In this case, the optimal path $\vect{x}(t)$ (i.e., the minimizer of $s\left[\vect{x}\left(t\right)\right]$) is along a straight line connecting the origin to 
$\vect{X}$, so the optimization problem is effectively one-dimensional. As a result
$P_{\text{st}}\left(\vect{X}\right) \sim P_{\text{st}}^{\left(1D\right)}\left(|\vect{X}|\right)$
where
$P_{\text{st}}^{\left(1D\right)}\left(X\right)$
is the SSD of a corresponding model in $d=1$ with potential $U\left(|x|\right)$ and rate function $\Phi(|z|)$.

One useful application is the RTP in $d=2$ whose speed is unity and which randomly reorients, at a constant rate $\tau^{-1}$ to a new orientation that is uniformly chosen from the unit circle. For this model, 
$P_{\text{free}}\left(\vect{x},t\right)$ was found exactly \cite{SBS20}, and it was shown that the LDP
\eqref{LDPfree}
 holds with 
$\Phi\left(\vect{z}\right)=1-\sqrt{1-z^{2}}$.
This rate function coincides with that of an RTP in $d=1$. As a result, 
so do the corresponding $\mu$'s, i.e., $\mu\left(u\right) = 2u/ \! \left(u^{2}-1\right)$.
For a harmonic confining potential, this prediction agrees with the recent exact result of \cite{Frydel22}, see \cite{SM}.

Another important application is to the ABP \cite{BechingerRev}, which involves an orientation vector that evolves via rotational diffusion. In $d=2$, the free ABP of unit speed is defined by the Langevin equations
\be
\dot{x}=\cos\theta\left(t\right), \quad \dot{y}=\sin\theta\left(t\right), \quad \dot{\theta}=\sqrt{2D_{r}} \, \eta\left(t\right),
\ee
where $D_r$ is the rotational diffusion coefficient and $\eta(t)$ is a white Gaussian noise with 
$\left\langle \eta\left(t\right)\right\rangle =0$
and
$\left\langle \eta\left(t\right)\eta\left(t'\right)\right\rangle =\delta\left(t-t'\right)$.
For the ABP, the LDP
$P_{\text{free}}\left(x,y,t\right)\sim e^{-tD_{r}\Phi\left(\sqrt{x^{2}+y^{2}}/t\right)}$
was shown (here $D_r$ plays the role of $\tau^{-1}$) and $\Phi(z)$ and $\lambda(k)$ were calculated exactly \cite{PKS16, Kurzthaler18, ABP2019}. The latter is given by 
$\lambda\left(k\right)=-a_{0}\left(-2k\right) / 4$
where $a_0(q)$ is the smallest eigenvalue of the Mathieu equation
\be
\label{Mathieu}
\psi''\left(v\right)+\left(a-2q\cos2v\right)\psi\left(v\right)=0
\ee
which admits a solution $\psi(v)$ of periodicity $\pi$.

For an ABP trapped in a central potential, 
our theory predicts that at large $D_r$,
$P_{\textrm{st}}\left(\vect{X}\right)\sim e^{-D_{r}s\left(X\right)}$ 
where
$s(X)$ is given by Eq.~\eqref{sXsol}. 
By numerically inverting $\lambda(k)$ while using the algorithm from \cite{PKS16} to compute $a_0(q)$, we have plotted $\mu(u)$ in Fig.~\ref{figMuOfu}.
The asymptotic behaviors of $a_0(q)$ are known in each of the limits $q \to 0$ and $q\to \infty$ \cite{Mathieu},
and from them we obtain in \cite{SM} the corresponding asymptotic behaviors
\be
\label{muOfuABP}
\mu\left(u\right)=\begin{cases}
-2u-\frac{7u^{3}}{2}+\dots, & \left|u\right|\ll1,\\[1mm]
\frac{-\text{sgn}\left(u\right)}{2\left(1-\left|u\right|\right)^{2}}+\frac{\text{sgn}\left(u\right)}{8\left(1-\left|u\right|\right)}+\dots,& 1-|u|\ll1.
\end{cases}
\ee
In \cite{SM} we show that the first line in Eq.~\eqref{muOfuABP} is in agreement with results of \cite{Malakar20} for an ABP in a harmonic trap.

We can immediately extend our approach to dynamics given by Eq.~\eqref{Langevin} where 
$\sigma = \sigma_1 + \dots + \sigma_n$
is the sum of $n$ statistically-independent noise terms, e.g., the sum of an active and a passive (thermal) noise.
In this case, it is straightforward to show that the SCGF $\lambda(k)$ is given by the sum of the SCGFs of the individual noises, i.e.,
 $\lambda(k) = \lambda_1(k) + \dots + \lambda_n(k)$.
 All that remains then is to find $\mu(u)$ [which involves inverting $-\lambda(k)/k$].
 In the particular case where $U(x)$ is a harmonic potential, the SSD is given by the convolution of the SSDs of the individual noises, due to the linearity of Eq.~\eqref{Langevin} \cite{SmithFarago22, SLMS22, Tucci22, Frydel22}.

Until now, we have tacitly assumed that the noise dynamics are not affected by 
$x(t)$.
We now relax this assumption. Consider first a stochastic process $\sigma(t)$ whose evolution depends on some time-dependent parameter $a(t)$ which can represent, e.g., the tumbling rate for an RTP, rotational diffusion coefficient for an ABP, etc. Let $\Phi_{a}\left(z\right)$ be the rate function that describes the position distribution for a free particle with the noise $\sigma(t)$ for constant $a(t) \equiv a$, and denote $\Phi_{a}$'s Legendre-Fenchel transform by $\lambda_a(k)$.
We now relate these functions to the SSD of a trapped active particle, evolving according to Eq.~\eqref{Langevin} where $\sigma$ evolves in time with $a(t) = x(t)$.
By slightly modifying the derivation presented above, we find that Eqs.~\eqref{action} and \eqref{sXsol} give way to
$s\left[x\left(t\right)\right]=\int_{-\infty}^{0}\Phi_{x\left(t\right)}\left(\dot{x}-F\left(x\right)\right)dt$
and
$s=\int_{0}^{X}p \, dx=\int_{0}^{X}\mu_{x}\left(F\left(x\right)\right)dx,$
respectively, where $\mu_{a}(u)$ is the inverse function of $-\lambda_a(k)/k$.
In particular, this extension allows one to treat multiplicative noise, including for instance the case in which the force $F(x)$ is intermittent, stochastically switching on and off \cite{MBMS20, SDN21, MBM22}.
For particular types of noises, this 
$s\left[x\left(t\right)\right]$  is of similar form as 
was found in other settings, such as urn models or non-Markovian random walks \cite{Franchini17, JH20, FB23}, or population dynamics \cite{OvMe2010, MeersonAssaf2017, SM}.

\textit{Discussion} --- We have calculated the SSD and MFPT for a generic trapped active particle in the limit $\tau \to 0$, uncovering a remarkable connection between the LDF $s(X)$ and the free rate function $\Phi(z)$. 
Our results are very general as they are valid for arbitrary active particle in an arbitrary potential in $d=1$ under very mild assumptions, and in $d>1$ under the additional assumption of rotational symmetry.

Generically, $\Phi\left(z\right)\simeq Az^{2}$ is parabolic around its minimum $z=0$, leading to a linear behavior $\mu\left(u\right)\simeq-4Au$ around $u=0$, see Fig.~\ref{figMuOfu}. Plugging this into \eqref{sXsol} provides a smooth matching with the Boltzmann distribution that describes typical fluctuations of $\vect{X}$, with  effective temperature $k_B T_\text{eff} = \tau / (4A)$.
Although we assumed here that the external potential is trapping, with a single minimum, our results for the MFPT yield transition rates between metastable states, e.g. for double-well \cite{CMPV19} or periodic \cite{led20, SH23} potentials.

The theoretical framework used here can be immediately extended to more general settings, e.g., beyond the overdamped limit as was recently demonstrated in Ref.~\cite{SmithFarago22} for the particular case of a harmonic trapping potential. One can also go beyond the steady state, and study dynamics of the position distribution, by  minimizing the action functional over coarse-grained histories defined on a finite time interval.

The OFM also yields the optimal history of the system conditioned on $X$. Moreover, by comparing Eqs.~\eqref{LangevinCG} and our definition $p=\Phi'\left(\dot{x}-F\left(x\right)\right)$, 
 we find that $\lambda'\left(p(t)\right)$ is the optimal realization of the noise $\bar{\sigma}(t)$ conditioned on $X$ (where we used that $\lambda'\left(k\right)$ is the inverse of the function $\Phi'\left(z\right)$ \cite{footnote:LegendreFenchel}).
 
 One can study other models of active noises $\sigma(t)$. One example is a shot noise which appears, for instance, in jump processes in which $x(t)$ takes only integer values. In this case, as we show in \cite{SM}, our theoretical framework reproduces the WKB theory that has been widely used in population dynamics 
 \cite{OvMe2010, MeersonAssaf2017}.


From our single-particle results it should be straightforward to deduce properties of a gas of noninteracting particles. For instance, the single-particle position distribution is proportional to the density of such a gas, but also other properties may be inferred such as first-passage \cite{RednerMeerson, AgranovMeerson18} and extreme-value statistics \cite{ExtremeValueStatisticsReview}.
It would be interesting to try to extend our results to other types of active processes, to multiple interacting active particles \cite{LMS21, Singh21, ARYL21, PBDN21, RSBI22, Cates22, ARYL22, KPS23}, and to disordered systems \cite{Woillez20}. 
Finally, our result \eqref{sXsol}, together with the relation between $\mu$ and $\Phi$, enables one to experimentally determine one of the three functions $s$, $\Phi$ and $F$ by measuring the other two.

\bigskip

{\it Acknowledgments ---} 
I acknowledge useful discussions with Baruch Meerson, Pierre Le Doussal and Oded Farago, and a helpful correspondence with Robert L. Jack.

{}

\newpage

\begin{widetext}
	\newpage

	\section*{Supplemental Material to the paper {``Nonequilibrium steady state of trapped active particles"} by N. R. Smith}

\renewcommand{\theequation}{S\arabic{equation}}
\renewcommand{\thefigure}{S\arabic{figure}}
\setcounter{equation}{0}
\setcounter{figure}{1}

\renewcommand*{\citenumfont}[1]{S#1}
\renewcommand*{\bibnumfmt}[1]{[S#1]}

\subsection{Symmetric tun-and-tumble particle in one dimension}

As described in the main text, a standard Run-and-tumble particle (RTP) $x(t)$ in dimension $d=1$, with an external force $F(x)$ is described by the equation
\be
\dot{x}=F\left(x\right)+\sigma\left(t\right)
\ee
where $\sigma(t)$ is telegraphic (dichotomous) noise of unit amplitude, $\sigma(t) = \pm 1$, that switches sign at a constant rate $\tau^{-1}$.
The steady-state distribution (SSD) of the position of such an RTP is given, up to a normalization constant, by
\be
\label{eq:Pst_2st}
P_{\textrm{st}}\left(X\right)\propto\frac{1}{1-F^{2}\left(X\right)}\exp\left[\frac{2}{\tau}\int_{0}^{X}dx\frac{F\left(x\right)}{1-F^{2}\left(x\right)}\right] \, .
\ee
The result \eqref{eq:Pst_2st} was obtained originally in the study of quantum optics, long ago \cite{q-optics1s,q-optics2s,q-optics3s,q-optics4s}, and afterwards reproduced in the context of dynamical systems with colored noise \cite{VBH84s, coloreds} and later, in the study of active matter \cite{Kardar2015s, Dhar_2019s}.
The leading-order $\tau \to 0$ behavior of Eq.~\eqref{eq:Pst_2st} is obtained by neglecting the pre-exponential factor (including the normalization constant), and one can rewrite the result in the form
\be
P_{\textrm{st}}\left(X\right)\sim e^{-\frac{1}{\tau}\int_{0}^{X}\mu\left(F\left(x\right)\right)dx},\qquad\mu\left(u\right)=\frac{2u}{u^{2}-1},
\ee
which is in perfect agreement with the prediction given in the main text, thus corroborating the validity of the theoretical method that we used.

%

\subsection{Asymmetric tun-and-tumble particle in one dimension}

Let us consider the asymmetric RTP as defined in the main text. Here $\sigma\left(t\right)\in\left\{ 1,-\alpha\right\} $, and the dynamics of the corresponding probability vector
$\left( \! \begin{array}{c}
p_{1}\\
p_{-\alpha}
\end{array} \! \right)$
is described by the master equation
\be
\frac{d}{dt}\left(\begin{array}{c}
p_{1}\\
p_{-\alpha}
\end{array}\right)=\frac{1}{\tau}L^{\dagger}\left(\begin{array}{c}
p_{1}\\
p_{-\alpha}
\end{array}\right),\qquad \text{where} \quad L^{\dagger}=\left(\begin{array}{cc}
-1 & \alpha\\
1 & -\alpha
\end{array}\right)
\ee
is the generator of the dynamics.
First of all, it is easy to see that the steady state of the noise is given by
$\left(\begin{array}{c}
p_{1}\\
p_{-\alpha}
\end{array}\right)=\frac{1}{1+\alpha}\left(\begin{array}{c}
\alpha\\
1
\end{array}\right)$, so that the noise has zero average $\left\langle \sigma\left(t\right)\right\rangle =0$, as noted in the main text.

The SCGF $\lambda(k)$ that corresponds to this asymmetric noise has been found in \cite{MeersonZilber18s}, but for the sake of completeness we will derive it here as well.
We use the Donsker-Varadhan formalism \cite{Bray, Majumdar2007s, DonskerVaradhans,Ellis,hugo2009,Touchette2018s}. The position of a free such RTP can be written as
\be
x\left(t\right)=\int_{0}^{t}\sigma\left(t\right)dt=\int_{0}^{t}\left[\delta_{\sigma\left(t\right),1}-\alpha\delta_{\sigma\left(t\right),-\alpha}\right]dt \, .
\ee
By the G\"{a}rtner-Ellis theorem \cite{Elliss}, $\lambda(k)$ is given by the largest eigenvalue of the ``tilted generator''
\be
L_{k}^{\dagger}=L^{\dagger}+k\left(\begin{array}{cc}
1 & 0\\
0 & -\alpha
\end{array}\right)\,.
\ee
By a direct calculation, this eigenvalue is found to be
\be
\lambda\left(k\right)=\frac{-\alpha\left(k+1\right)+\sqrt{\left(\alpha+1\right)\left[\alpha\left(k+1\right)^{2}+\left(k-1\right)^{2}\right]}+k-1}{2},
\ee
coinciding with Eq.~\eqref{lambdakAsymmtricRTP} of the main text.
The corresponding rate function $\Phi(z)$ can be found by taking the Legendre-Fenchel transform of $\lambda(k)$ \cite{MeersonZilber18s}, but this is not needed for our purposes of calculating the position SSD for a trapped particle.

\subsection{Pearson random walk (PRW)}

Let us first consider a generalization of the PRW as defined in the main text. Namely, we will assume that tumbling events take place at discrete times $t=\tau, 2\tau, 3\tau, \dots$, $\sigma$. At each tumbling event, $\sigma$ is set to a value chosen from some given distribution $p(v)$, and remains constant until (possibly) changing at the next tumbling event.
For the particular case
$p\left(v\right)=\frac{\delta\left(v-1\right)+\delta\left(v+1\right)}{2}$
we recover the PRW as defined in the main text.

We consider first a free particle. Then it is easy to see that
$x\left(t=n\tau\right)=\tau\sum_{i=1}^{n}\sigma_{i}$,
where $\sigma_i$ is the value of $\sigma(t)$ in the time interval
$t\in\left[\left(i-1\right)\tau,i\tau\right]$.
Now, the $\sigma_i$'s are independent and identically distributed (i.i.d.) random variables, so that the generating function of the distribution $P_{\text{free}}\left(x,t\right)$
\be
\int_{-\infty}^{\infty}e^{kx/\tau}P_{\text{free}}\left(x,t\right)dx=\left[\int_{-\infty}^{\infty}e^{kv}p\left(v\right)dv\right]^{n} \, .
\ee
Now, using that
$\int_{-\infty}^{\infty}e^{kx/\tau}P_{\text{free}}\left(x,t\right)dx\sim e^{t\lambda\left(k\right)/\tau}$
with $n = t/\tau$, we find that
\be
\label{lambdakGeneralPRW}
\lambda\left(k\right)=\ln\int_{-\infty}^{\infty}e^{kv}p\left(v\right)dv \, .
\ee

From here onwards we focus on the case 
$p\left(v\right)=\frac{\delta\left(v-1\right)+\delta\left(v+1\right)}{2}$,
corresponding to the standard PRW.
In this case, Eq.~\eqref{lambdakGeneralPRW} becomes
\be
\lambda\left(k\right)=\ln\left(\cosh k\right) \, .
\ee
The Legendre-Fenchel transform of this function gives the well-known rate function \cite{MS2017s}
\be
\Phi(z) = \frac{\left(1+z\right)\ln\left(1+z\right)+\left(1-z\right)\ln\left(1-z\right)}{2 }\, .
\ee
Note that this rate function also describes a random walk in discrete time and space (because a free PRW in $d=1$ can clearly be viewed as such, if one considers only the position at times $0, \tau, 2\tau, \dots$). Such random walks have been studied using an approach that is very similar to our coarse graining in \cite{Harris15s, Jack19s, JH20s}, where this approach was called the ``temporal additivity principle'' \cite{HT09s}.
The function $\mu(u)$, which is the inverse of the function $-\lambda(k)/k$, can be obtained numerically. Its asymptotic behaviors are not difficult to calculate. Using
\be
\lambda\left(k\right)=\begin{cases}
-\frac{k}{2}+\frac{k^{3}}{12}+\dots, & k\to0,\\[2mm]
-1+\frac{\ln2}{\left|k\right|}+\dots & \left|k\right|\to\infty,
\end{cases}
\ee
we find that
\be
\label{muOfuPRWSM}
\mu\left(u\right)=\begin{cases}
-2u-\frac{4u^{3}}{3}+\dots, & u\to0,\\[2mm]
-\text{sgn}\left(u\right)\frac{\ln2}{1-\left|u\right|}+\dots, & |u|\to1,
\end{cases}
\ee
which is Eq.~\eqref{muOfuPRW} in the main text.
The exact $\mu(u)$ is plotted together with its asymptotic behaviors \eqref{muOfuPRWSM}, in Fig.~\ref{figMuOfuABP}.

\subsection{RTP in $d=2$: Comparison with Frydel \cite{Frydel22s}}

One of the models studied by Frydel in \cite{Frydel22s} was an RTP in $d=2$, in which the orientation $\vect{\sigma}(t)$ switches at rate $\tau^{-1}$ to a new orientation chosen uniformly from the unit circle, trapped by a harmonic potential $U(\vect{x}) = kx^2 / 2$.
He found that the exact SSD is given (up to a normalization constant) is given by
\be
\label{PstRTP2Dexact}
P_{\text{st}}\left(\vect{X}\right)\propto\left(1-k^2 X^{2}\right)^{\alpha-1},
\ee
where
$\alpha = 1/(k\tau)$.
Indeed, using our general result \eqref{sXsol} of the main text, with
\be
F\left(x\right)=-kx,\qquad\mu\left(u\right)=\frac{2u}{u^{2}-1}
\ee
we find
\be
P_{\text{st}}\left(\vect{X}\right)\sim e^{-s\left(X\right)/\tau}=\left(1-k^{2}X^{2}\right)^{1/\left(k\tau\right)} \, ,
\ee
which agrees with Eq.~\eqref{PstRTP2Dexact} in the leading order, in the small-$\tau$ limit.

\begin{figure}[t]
\includegraphics[width=0.45\linewidth,clip=]{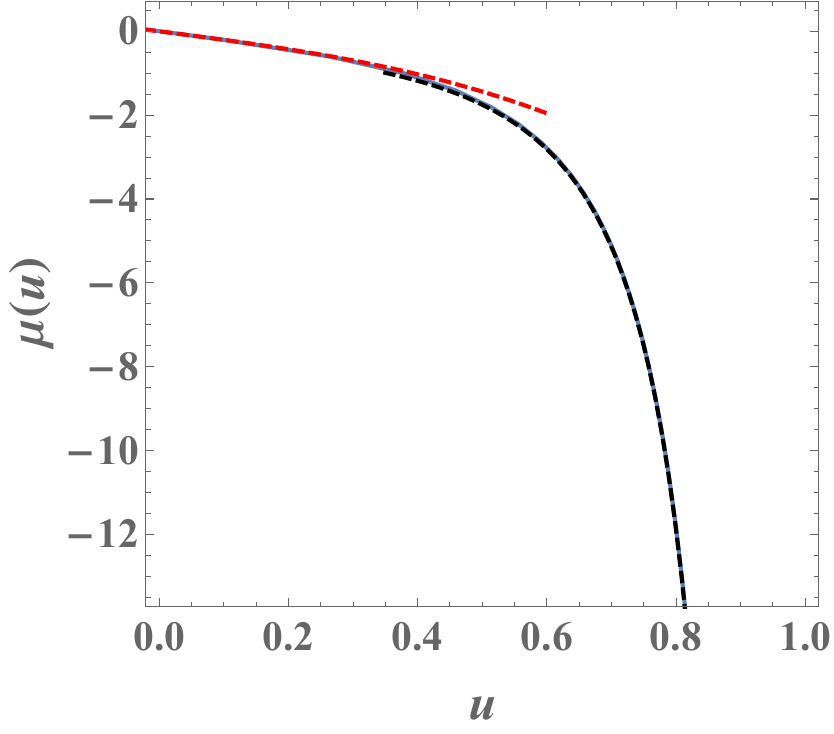}
\includegraphics[width=0.45\linewidth,clip=]{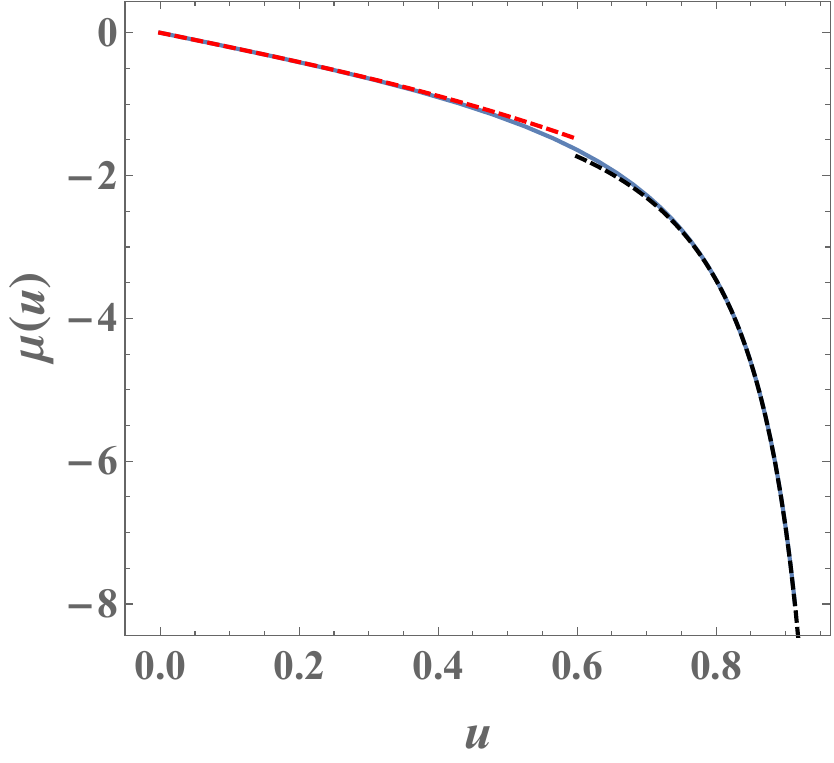}
\caption{(a) The function $\mu(u)$ for the PRW (solid line) together with its asymptotic behaviors given in Eq.~\eqref{muOfuPRWSM} (dashed lines). Due to the relation $\mu(u) = - \mu(-u)$, only the region $u>0$ is plotted. (b) Similarly for the ABP, where the asymptotic behaviors are given in Eq.~\eqref{muOfuABP} in the main text.}
\label{figMuOfuABP}
\end{figure}

\subsection{Asymptotic behaviors for the ABP}

The behaviors of $a_0(q)$  for $q \to 0$ and $q\to \infty$ take the forms
\be
a_{0}\left(q\right)=\begin{cases}
\sum_{n=1}\alpha_{2n}\,q^{2n} & \text{for}\;\;q\to0,\\[2mm]
\sum_{n=0}\beta_{n}\,q^{1-\frac{n}{2}} & \text{for}\;\;q\to\infty\;.
\end{cases}
\ee
Explicit values of the coefficients $\alpha_{2n}$ and $\beta_n$ are known \cite{Mathieus} and the first few are given by
\bea
\label{eq:alpha_n}
&& \alpha_2 = -\frac 12,\qquad \alpha_4 = \frac 7{128}, \qquad \alpha_6 = - \frac{29}{2304}, \qquad \alpha_8 = \frac{68687}{18874368}, \;\; \cdots , \\[1mm]
\label{eq:beta_n}
&& \beta_0 = -2, \qquad \beta_1 = 2, \qquad \beta_2 = -\frac{1}{4},\qquad
\beta_3 = -\frac 1{32}, \qquad \beta_4 = - \frac 3{256}, \qquad
\beta_5=-\frac {53}{8192},\; \cdots .
\eea
We now use these behaviors in order to derive corresponding behaviors of $\mu(u)$ for $u\to0$ and $|u| \to 1$, respectively. First of all, using the leading two terms in the expansion of $a_0(q)$ for $q \to 0$ together with $\lambda\left(k\right)=-\frac{1}{4}a_{0}\left(-2k\right)$, we obtain
\be
\lambda\left(k\right)=\frac{k^{2}}{2}-\frac{7k^{4}}{32}+\dots, \qquad k\to0,
\ee
and therefore, in the leading order, the inverse function of 
$-\lambda\left(k\right)/k=-\frac{k}{2}+\frac{7k^{3}}{32}+\dots$
is
\be
\label{muOfuABPSmall}
\mu\left(u\right) = -2u-\frac{7u^{3}}{2}+\dots, \qquad u \to 0.
\ee
In the opposite limit, by using the leading three terms in the expansion of $a_0(q)$ for $q \to \infty$, we obtain
\be
\lambda\left(k\right)=k-\sqrt{\frac{k}{2}}+\frac{1}{16}+\dots, \qquad k \to \infty.
\ee
Then the inverse function of $-\lambda(k) / k$ behaves as
\be
\label{muOfuABPLarge}
\mu\left(u\right)=\frac{1}{2\left(u+1\right)^{2}}-\frac{1}{8\left(u+1\right)}+\dots, \qquad u \to -1.
\ee
Eqs.~\eqref{muOfuABPSmall} and \eqref{muOfuABPLarge} together yield Eq.~\eqref{muOfuABP} in the main text, where we also used that, due to the mirror symmetry of the noise, $\mu(u) = - \mu(-u)$ is an odd function of $u$.
The exact $\mu(u)$, which we calculated by using the numerical algorithm given in \cite{PKS16s} for computing $a_0(q)$, is plotted together with its asymptotic behaviors in Fig.~\ref{figMuOfuABP}.

Let us now consider the particular case of an ABP trapped in a harmonic potential $U(\vect{x}) = x^2 /2$, for which we can compare some of our predictions with existing results. In \cite{Malakar20s}, an exact expression was obtained for the SSD of an ABP, which also experiences translational diffusion, trapped in a harmonic potential. The SSD is given in the form of a rather complicated infinite series. In the limit of zero translational diffusion and large rotational diffusion, this expression simplifies significantly, as is given in Eq.~(15) of \cite{Malakar20s} which, using our notation and choice of units, reads
\be
\label{ABPFromMalakar}
P\left(\vect{X}\right)=2D_{r}e^{-D_{r}X^{2}}\left[1-\frac{1}{D_{r}}\left(\frac{3}{4}-\frac{5}{2}D_{r}X^{2}+\frac{7}{8}D_{r}^{2}X^{4}\right)+O\left(\frac{1}{D_{r}^{2}}\right)\right].
\ee
Eq.~\eqref{ABPFromMalakar} is valid in the limit $D_r \to \infty$ at constant $D_r X^2$.

We now compare this to the prediction that comes from our results. Using Eq.~\eqref{muOfuABPSmall} we find that, for a harmonic potential $U(\vect{x}) = x^2 /2$,
\be
s\left(X \ll 1\right)=\int_{0}^{X}\mu\left(F\left(x\right)\right)dx=\int_{0}^{X}\mu\left(-x\right)dx\simeq\int_{0}^{X}\left(-2\left(-x\right)-\frac{7}{2}\left(-x\right)^{3}\right)dx=X^{2}+\frac{7}{8}X^{4} \, ,
\ee
so the SSD is given by
\be
\label{ABPsmallX}
P_{\textrm{st}}\left(X\right)\sim e^{-D_{r}\left(X^{2}+7X^{4}/8\right)} \, .
\ee
Eq.~\eqref{ABPsmallX} holds in the limit $X\ll 1 \ll D_r$. 

In order to compare our result with the result of \cite{Malakar20s}, we now analyze both of them in the limit $D_{r}X^{4}\ll1\ll D_{r}X^{2}$ in which they are both expected to hold. 
In this limit, Eq.~\eqref{ABPFromMalakar} simplifies to
\be
\label{ABPFromMalakar2}
P\left(\vect{X}\right) \simeq 2D_{r}e^{-D_{r}X^{2}}\left(1-\frac{7}{8}D_{r}X^{4}\right).
\ee
while Eq.~\eqref{ABPsmallX} becomes
\be
P_{\textrm{st}}\left(X\right)\sim e^{-D_{r}X^{2}}e^{-7D_{r}X^{4}/8}\simeq e^{-D_{r}X^{2}}\left(1-D_{r}\frac{7}{8}X^{4}\right)\,.
\ee
The two predictions indeed agree, up to the normalization factor which is beyond the accuracy of our leading-order large-deviation calculations.


\subsection{Shot noise}

Let $\sigma(t)$ be a shot noise, such that
\be
\label{shotNoiseDef}
\sigma\left(t\right)=\tau\sum_{i}\left(t-t_{i}\right)
\ee
where $t_1,t_2,\dots$ is a Poisson process with density $\tau^{-1}$.
For a free particle evolving according to
$\dot{x}\left(t\right)=\sigma\left(t\right)$,
$x(t)$ takes discrete values so it is in fact a jump process. Such processes have been studied in many different contexts, including in particular stochastic population dynamics in which $x(t)$ would correspond to the population size at time $t$. A large-deviations approach based on a discrete WKB approximation to the solution of the master equation has been formulated quite some time ago and employed in many different ecological systems \cite{OvMe2010s, MeersonAssaf2017s}.
Here we will show that our approach described in the main text, namely that of coarse-graining the dynamics by applying the LDP on time windows of intermediate sizes, gives rise to the same formalism as the discrete WKB approach.

Let us first find the rate function $\Phi(z)$.
For this we first note that $x(t)/\tau$ is a Poisson random variable with mean $t/\tau$:
\be
P\left(x/\tau=n\right)=\frac{\left(t/\tau\right)^{n}e^{-t/\tau}}{n!}.
\ee
Using Stirling's approximation, $\ln n!\simeq n\ln n-n$, we find that at $n\gg1$,
\be
\ln P\left(x/\tau=n\right)\simeq\frac{t}{\tau}\left(\frac{n\tau}{t}\ln\frac{t}{n\tau}+\frac{n\tau}{t}-1\right) \, ,
\ee
from which one finds that $x(t)$ obeys a large-deviations principle (LDP) 
$P_{\text{free}}\left(x,t\right)\sim e^{-(t/\tau)\Phi\left(x/t\right)}$
with rate function
\be
\Phi\left(z\right)=z\ln z-z+1\,,
\ee
whose Legendre-Fenchel transform is
\be
\lambda\left(k\right)=e^{k}-1 \, .
\ee
Therefore, for a free such particle, the Hamiltonian corresponding to the action functional $s[x(t)]$ is simply
$H = e^{p}-1$,
coinciding with the Hamiltonian that one obtains from the large-deviation formalism that follows from the discrete WKB approximation \cite{OvMe2010s, MeersonAssaf2017s}.

Note that our approach enables one to extend the analysis, for instance (as described in the main text) to the case where $\sigma(t)$ is the sum of a shot noise and some other noise (e.g., white noise), which would be difficult from the point of view of the discrete WKB approximation \cite{OvMe2010s, MeersonAssaf2017s}, as the latter relies on the discreteness of $x(t)$.

\end{widetext}


\begin{thebibliography}{}
\bibitem{footnote:drift} If $\vect{z}^* \ne 0$, one can study instead the distribution of 
$\vect{x}\left(t\right)-\left\langle \vect{x}\left(t\right)\right\rangle $.
\bibitem{footnote:smallTau} Note that when taking the limit $\tau\to0$, we do not simultaneously take the amplitude of the noise $\sigma$ to infinity, so we do not recover the white noise limit as $\tau \to 0$, cf. \cite{SmithFarago22}.
\bibitem{footnote:LegendreFenchel} For simplicity, we will assume that the functions $\Phi(z)$ and $\lambda(k)$ are both strictly convex and differentiable, so that their Legendre-Fenchel transforms reduce to the Legendre transforms \cite{TouchetteNutshell}, i.e., $\lambda(k)$ is obtained from $\Phi(z)$ via
$k=\Phi'\left(z\right),\quad\lambda=kz-\Phi$ and vice versa.
\end{thebibliography}
\end{document}